\documentclass{appolb}
\usepackage{epsfig}

\def\la{\langle}
\def\ra{\rangle}

\begin{document}
\title{Transport of Forced Quantum Motors in the Strong Friction
Limit
\thanks{Presented at the Marian Smoluchowski Symposium on Statistical Physics, Krak\'ow, May 14-17, 2006}
}
\author{Lukasz Machura, Jerzy {\L}uczka
\address{Institute of Physics, University of Silesia,
    40-007 Katowice, Poland}
\and Peter Talkner, Peter H\"anggi
\address{Institut f\"ur Physik, Universit\"at Augsburg,
 86135 Augsburg, Germany}
}
\maketitle
\begin{abstract}
The directed transport of an overdamped Brownian
motor moving in a spatially periodic  potential that lacks
reflection symmetry (i.e. a ratchet potential) is studied when
driven by thermal and dichotomic nonequilibrium noise in the
presence of an external, constant load force. We consider both, the
classical and the  quantum tunneling assisted regimes. The
current-load characteristics are investigated as a function of the
system parameters like the load force, the temperature and the
amplitude strength of the applied  two-state noise.
\end{abstract}

\section{Introduction}
Classical regimes of transport of microscopic objects like
Brown\-ian particles are well elaborated in the previous literature
(for a historical overview see in Ref. \cite{chaos}). In the last
decade, special interest has been devoted to transport in  ratchet
systems (also termed  Brownian motor systems), i.e. to the
phenomenon of noise assisted, directed motion of particles in
spatially periodic structures which possess a broken reflection
symmetry \cite{clas1,clas2,clas3}. In contrast, the quantum
properties of directed transport are only partially elaborated in
such Brownian motor systems
\cite{quant1,quant2,MacKos2004,MacKos2006}. Challenges arise in the
quantum regime because the transport can strongly depend on the
mutual interplay of pure quantum effects like tunneling and particle
wave interference with  dissipation processes,  nonequilibrium
fluctuations and external driving \cite{ingoldchaos}. Moreover,
there exist typically no analogous closed evolution equations of
such system as in the classical regimes, which are based for example
on Langevin or Fokker-Planck equations. However, special quantum
regimes can be described nevertheless by use of  effectively
classical methods. For example, if the system strongly interacts
with a thermostat, quantum diffusive dynamics can be described by an
effective classical Smoluchowski equation for the diagonal part of
the statistical operator in the position representation
\cite{anker}, in which the potential and diffusion coefficient are
modified due to quantum effects (section 2). This so called quantum
Smoluchowski equation has been applied to describe activation
processes, quantum diffusion, and Brownian motors
\cite{quant1,quant2,MacKos2004,MacKos2006,anker}. In this work, we
employ it to study the transport properties of an overdamped
Brownian motor moving in a spatially periodic potential
$U(x)=U(x+L)$ of the period $L$ under the influence of an external,
constant  bias force when driven by both, thermal equilibrium and
nonequilibrium fluctuations. We analyze the classical as well as the
quantum regimes. In particular, the resulting {\em current-load
characteristics} are investigated as functions of the system
parameters like load, temperature and amplitude of the
nonequilibrium noise (section 3).

\section{Quantum Smoluchowski Equation}

For  systems strongly interacting with a thermostat, which in turn
implies a   strong friction limit, the quantum dynamics above the
crossover temperature to pure quantum tunneling \cite{ingold85} can
be described in terms of a generalized Smoluchowski equation which
accounts for the leading quantum corrections. For a particle of mass
$M$ moving in the potential $V(x)$, this quantum Smoluchowski
equation (QSE) for the coordinate-diagonal elements of the density operator
$\rho(t)$, i.e. for  the probability density function 
$P(x,t) = \langle x\vert \rho(t) \vert x\rangle$ in position space $x$, 
takes the form \cite{anker}:
\begin{equation} \label{S1}
\Gamma \frac{\partial}{\partial t} P(x,t) =
\frac{\partial }{\partial x} V'_\textrm{eff}(x) P(x,t)
+  \frac{\partial^2}{\partial x^2} D_\textrm{eff}(x) P(x, t),
\end{equation}
where $\Gamma$ denotes the friction coefficient. The effective potential reads
\begin{equation} \label{effpot}
V_\textrm{eff}(x) = V(x) + \frac{1}{2} \lambda V''(x),
\end{equation}
where the prime denotes the derivative with respect to the coordinate $x$.
The quantum correction parameter $\lambda$ describes  quantum fluctuations in  position space and reads
\begin{equation}\label{Lam1}
\lambda = \frac{ \hbar }{\pi \Gamma }\left[ \gamma +
\Psi\left(1+\frac{\hbar \beta \Gamma}{2\pi M}\right)  \right],
\quad \beta = \frac{1}{k_B T}.
\end{equation}
Here, $\Psi(z)$ is the digamma function,
$\gamma \simeq 0.5772$  the Euler-Mascheroni constant, $T$ is the
temperature and $k_B$ denotes the Boltzmann constant. The
parameter $\lambda$ depends nonlinearly on the Planck constant
$\hbar$ and on the mass $M$ of the Brownian particle (let us remind
that in the classical case, the overdamped dynamics does not depend
on the mass $M$).

 The effective diffusion coefficient reads
\cite{MacKos2004,LucRud2005}
\begin{eqnarray} \label{D1}
  D_\textrm{eff}(x) = \frac{1}{\beta [1 - \lambda \beta V''(x)]}.
\end{eqnarray}
Note that  for $k_BT \ll \hbar \Gamma/M$, $\lambda$ becomes
\begin{equation} \label{lam}
  \lambda = \frac{ \hbar }{\pi \Gamma }
  \left[\gamma+\ln \Big(\frac{\hbar \beta \Gamma}{2\pi M}\Big)\right].
\end{equation}
From the mathematical point of view, the  Smoluchowski 
equation (\ref{S1}) corresponds to the 
 classical Langevin equation in the Ito interpretation \cite{HT},
\begin{equation}\label{LE}
\Gamma \dot{x} = - V_\textrm{eff}'(x)  + \sqrt{2\Gamma
D_\textrm{eff}(x)}~\xi(t) \;.
\end{equation}
The zero-mean and the $\delta$-correlated Gaussian white noise
$\xi(t)$, meaning that  $<\xi(t)\xi(s)>=  \delta(t-s)$, models the
influence of a thermostat of temperature $T$ on the system.

\section{Biased Quantum Motor Transport}
We focus on the dynamics of overdamped quantum Brownian motors
\cite{quant1,quant2,MacKos2004,MacKos2006} moving in a  spatially 
periodic potential $U(x)=U(x+L)$ and  driven by nonequilibrium
fluctuations $\eta(t)$. The quantum thermal fluctuations are
determined by the parameter $\lambda$ (see
Eq. \ref{Lam1}). Additionally, a constant bias force $F_0$ is applied 
to the system. 
The dynamics can then be described by the Langevin equation
\begin{equation} \label{Lan}
\Gamma  \dot{x} = -V'_\textrm{eff}(x)  + \sqrt{2\Gamma
  D_\textrm{eff}(x)} \; \xi(t)
+ \eta(t) \;, 
\end{equation}
where $V_\textrm{eff}(x)$ is given by Eq. (\ref{effpot}) with 
\begin{equation} \label{potF}
V(x) = U(x)-F_0x.  
\end{equation}
We rewrite Eq. (\ref{Lan}) in the dimensionless form, namely, 
\begin{eqnarray}  \label{aa}
  \dot y = -W'_\textrm{eff}(y) + F + \sqrt{2 \mathcal{D}_\textrm{eff}(y)} \; \hat \xi (s)
+ \hat\eta(s),
\end{eqnarray}
where the position of the Brownian motor  is scaled as $y= x/L$,
time is rescaled as $s=t/\tau_0$, with the characteristic time
scale reading $\tau_0 = \Gamma L^2/\Delta  V$ (the barrier height
$\Delta V$ is the difference between the maximal and minimal values
of the unbiased potential $V(x)$). During this time span, a classical,
overdamped particle moves a distance of length $L$ under the
influence of the constant force $\Delta V/L$. The  effective
potential is $W_\textrm{eff}(y) = W(y) + (1/2) \lambda_0 W''(y)$,
where the rescaled periodic potential $W(y) =  U(yL)/ \Delta V = W(y+1)$
possesses unit period  and a unit barrier height. The dimensionless
parameter $\lambda_0 = \lambda /L^2$ describes  quantum fluctuations
over the characteristic length L, see in Ref. \cite{MacKos2004} for
further  details.

The rescaled diffusion function $\mathcal{D}_\textrm{eff} (y)$ reads,
\begin{eqnarray} \label{DD3}
  \mathcal{D}_\textrm{eff}(y) =\frac{1}{\beta_0[1 - \lambda_0 \beta_0  W''(y)]}\;.
\end{eqnarray}
The dimensionless, inverse temperature  $\beta_0 = \Delta V/k_B T$
is the ratio of the activation energy in the non-scaled potential
and the thermal energy. The rescaled Gaussian white noise reads
$\hat\xi(s)= ( L/\Delta V) \xi(t)$,
 the rescaled, nonthermal stochastic force is  $\hat\eta(s) =
 (L/\Delta V) \eta(t)$ and
the rescaled constant force stands for $F = (L/\Delta V) F_0$.

The  nonequilibrium fluctuations  $\hat\eta(s)$
in Eq. (\ref{aa}) are described by symmetric Markovian dichotomic noise
\begin{equation}
  \hat\eta(s) = \{-a,a\},\label{dichnois}
\end{equation}
which jumps between two states $a$ and $-a$ with a rate
$\nu$. The induced stationary probability current $J$, or equivalently the
asymptotic average velocity of the Brownian motor can then be determined
in the adiabatic limit in a closed form. Put differently, the above stated
problem can be  solved analytically in the limit 
$\nu \to 0$. In the above introduced dimensionless variables the
probability current takes the form
\begin{eqnarray}
  \la \dot{y} \ra &=& J = \frac{1}{2}\left[ J(a) + J(-a) \right], \\
  J(a) &=&  \frac{1-\exp [- \beta_0 (F+a)]} {\int\limits_0^1
 dy \; \mathcal{D}_\textrm{eff}^{-1}(y) \exp [-\beta_0\Phi(y,a)]
\int\limits_y^{y+1} dz \; \exp [ \beta_0\Phi(z,a)]}
\end{eqnarray}
with the biased, generalized thermodynamic  potential reading
\begin{eqnarray}
\Phi(y,a) &=& \int \frac{W'(y) - (F+a)y}{\mathcal{D}_\textrm{eff}(y)}\;dy 
\nonumber\\
&=& W(y) + (1/2) \lambda_0 W''(y) -(1/2) \lambda_0 \beta_0 [W'(y)]^2 
\nonumber\\
&-&  (1/4)\lambda^2_0 \beta_0 [W''(y)]^2 + (F+a) \lambda_0 \beta_0
 W'(y) - (F+a)y. 
\end{eqnarray}
Unbiased transport properties driven by such dichotomic noise
have been elaborated for classical particles in Ref. \cite{kul,KosLuc2001} and
in the quantum regime in Ref. \cite{MacKos2004}.

This analytic  expression for the current allows one  to study
directed transport in {\it arbitrarily shaped} ratchet potentials.
As an example, we consider a family of asymmetric periodic
potentials of the form
\begin{eqnarray}  \label{W(x)}
        W(y)= W_0 \{\sin(2\pi y)+0.4 \sin[4\pi(y-0.45)] \nonumber\\
     + B \sin [6\pi (y-0.45)]\},
\end{eqnarray}
where $B$ denotes a shape parameter and $W_0$
is chosen in such a way that the 
maximal variation of the potential is normalized to unity.

\begin{figure}[htbp]
  \begin{center}
    \includegraphics[angle=0,width=0.45\linewidth,clip=]{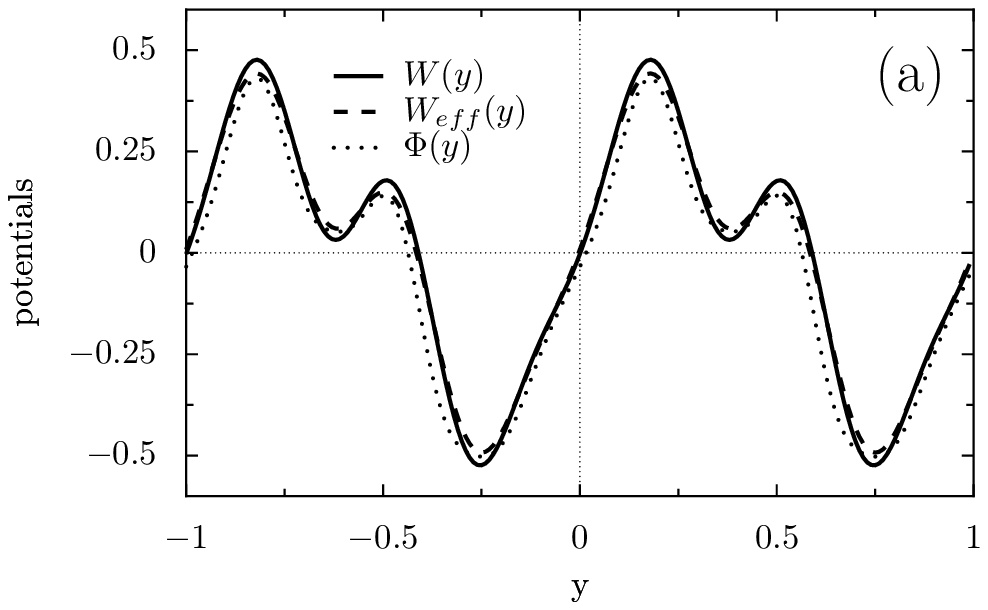}
    \includegraphics[angle=0,width=0.45\linewidth,clip=]{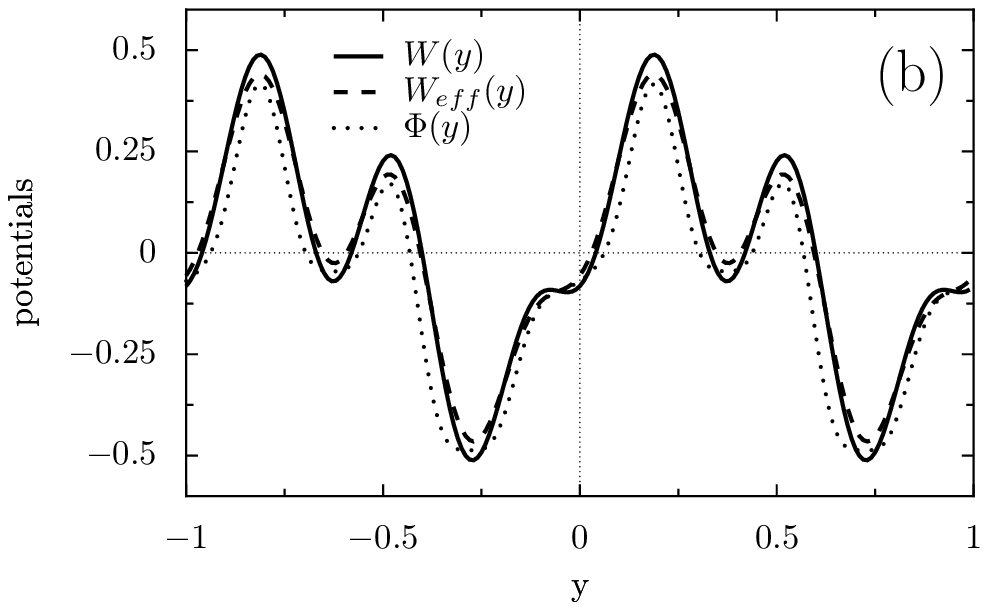}
  \end{center}
  \caption{The unbiased classical ratchet potential $W(y)$ (solid line)
    together with the corresponding unbiased 
  quantum potential $W_\textrm{eff}(y)$ (dashed line) and the generalized 
thermodynamic potential $\Phi(y)\equiv \Phi(y, 0)$ (dotted line) 
are depicted as
  functions of the scaled position 
    $y$ for $F=0$, $\beta_0=10$ and  two values of the shape parameter $B$.
The left panel (a) refers to $B=0.3$  and the panel (b) to
$B=0.62$. 
}
  \label{fig0}
\end{figure}

The influence of quantum corrections on the potential shape is
displayed in Fig. \ref{fig0} and on  transport of the biased
Brownian motor is presented in Figs. \ref{fig1}--\ref{fig3}. For the
rescaled quantum fluctuations, we set the temperature-dependent
quantum parameter (\ref{Lam1}) equal to $\lambda_0 = 10^{-4}[\gamma +
\Psi(1+10^4\beta_0)]$. We study the induced current-load
characteristics as a function of all system parameters and elucidate
how one can control the directed transport by adjusting the
(inverse) temperature $\beta_0$, the dichotomic noise strength $a$
and the strength of the constant force $F$.

\section{Quantum Transport  Characteristics}

\begin{figure}[htbp]
  \begin{center}
    \includegraphics[angle=0,width=0.9\linewidth,clip=]{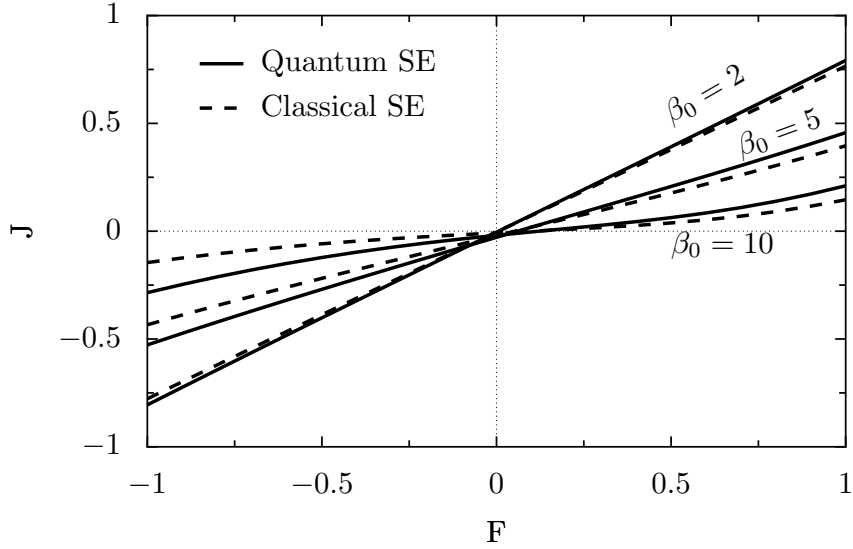}
  \end{center}
  \caption{(color online) The directed quantum noise-induced transport $J$ of
  the quantum Brownian motor (solid line) versus the constant bias
  force $F$ is compared with its
  classical limit (dashed line).  The current-load characteristics is studied here for
 several values of the dimensionless inverse temperature
  $\beta_0 = 2,5,10$.
  The dichotomic noise level is set to
  $a=1$. The ratchet potential is defined by $B=0.3$ (see panel (a)
  in Fig \ref{fig0}).
  }
  \label{fig1}
\end{figure}

We next address  the question of how the constant bias load $F$
affects the directed transport properties of the quantum and the
corresponding  classical Brownian motors that are driven out of
equilibrium by a nonthermal dichotomic random force $\eta(t)$. In
Fig. \ref{fig1} we present the current--load characteristics for the
dichotomic noise level $a=1.0$ and a ratchet potential with $B=0.3$,
see panel (a) in Fig. \ref{fig0}. The three sets of curves
correspond to three different values of the dimensionless inverse
temperature $\beta_0 = 2,5,10$.

\begin{figure}[htbp]
  \begin{center}
    \includegraphics[angle=0,width=0.9\linewidth,clip=]{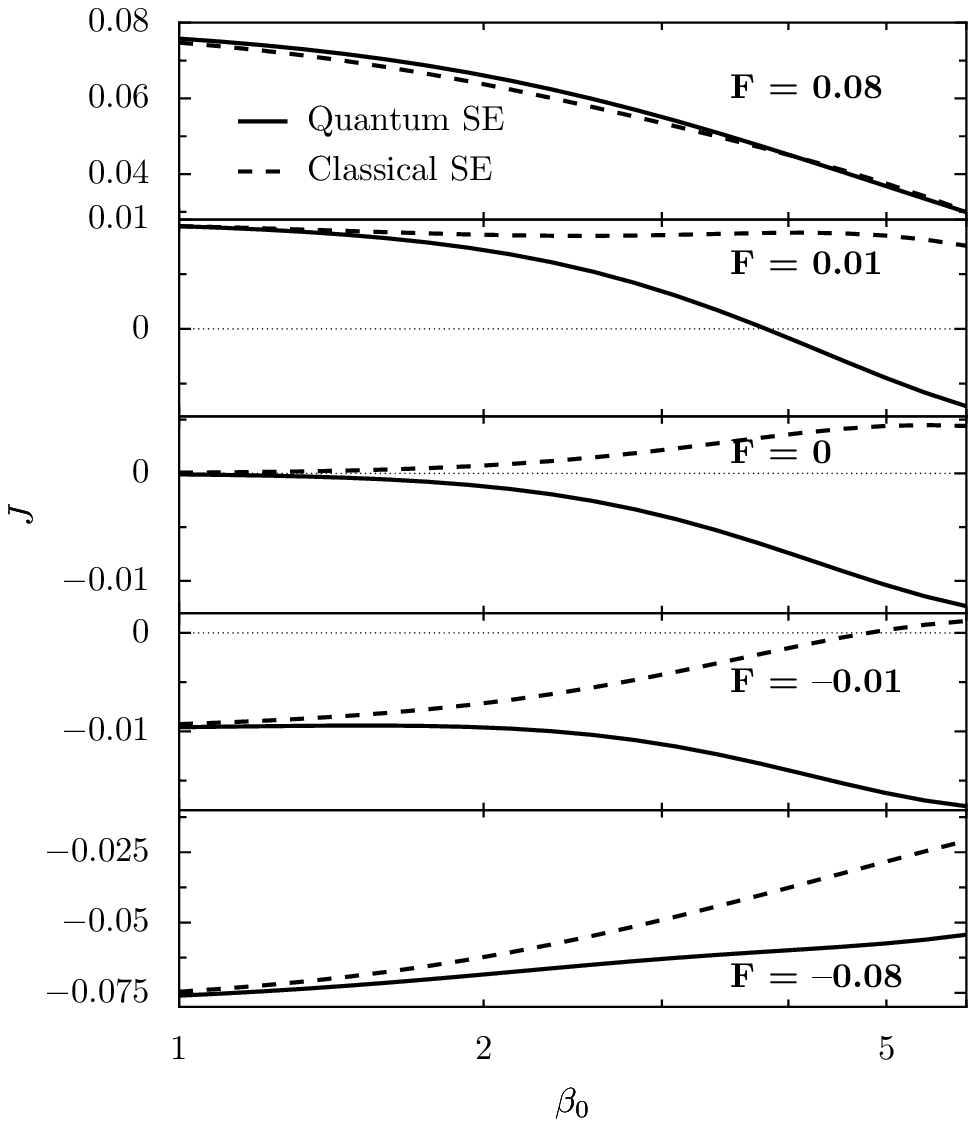}
  \end{center}
  \caption{The directed quantum noise-induced transport $J$ of
  the quantum Brownian motor (solid line) versus the dimensionless
  inverse temperature $\beta_0$ is compared with its
  classical limit (dashed line). The classical and quantum
  currents are depicted in the various panels  for
  five different values of the applied external bias $F$;
  i.e. for bottom to top $F=-0.08-0.01,0,0.01,0.08$.
  The dichotomic noise level is set to
  $a=1$. The ratchet potential is defined by $B=0.62$ (see panel (b)
  in Fig \ref{fig0}).
  }
  \label{fig2}
\end{figure}

In the absence of thermal (Gaussian) noise, the  dynamics of the
driven particle is confined to a single period as long as the bias
forces remain limited to  the interval [$F_1,F_2]$ with
$F_1 \simeq -3.6$ and $F_2 \simeq 4.42$ denoting the two threshold
values. Then, the dichotomic noise alone with $a=1$  is not able
to induce transitions to the neighboring periods; this  becomes
possible only in the presence of additional, thermal Gaussian noise
of unbounded amplitude which in turn induces  a finite probability
current. For larger thermal noise strength 
(i.e. for smaller $\beta_0$ or higher temperature), the quantum
corrections seemingly play only a minor role for the probability
current, see in Fig.~\ref{fig2}. It is only for lower temperatures
$T$ that the influence of quantum effects become more pronounced and
distinct deviations from the classical response behavior become
detectable.

The value of the constant bias, for which the current vanishes, is
termed the {\it stall force} $F_{stall}$. Generally, this stall
force depends on the temperature $\beta_0^{-1}$ and on the other
system parameters as well.
Fig. \ref{fig1} depicts that by cooling down the system (i.e.
increasing  the inverse temperature $\beta_0$) the stall force
becomes shifted toward larger  positive loads. This  means that for
lower temperatures the ratchet effect becomes  more pronounced.
Moreover, only for small enough temperatures one can resolve  the
different values of the stall force for the classical motor and the
quantum motor dynamics. If the temperature is high,
then both the quantum and the classical characteristics are very
similar and, additionally, both current-load characteristics cross
the zero-current axis  at values that are close to zero. This
corresponds to a rather  weak ratchet effect.

With a value for the two--state noise fluctuations set at  $a=1.0$
and for the ratchet potential defined by $B=0.62$ depicted  in panel
(b) in Fig. \ref{fig0}, we observe a pronounced influence of the
quantum corrections on the transport \cite{MacKos2004}. Also in this
case with $B=0.62$ the dichotomic force amplitude $\eta(t)$ alone
cannot induce transport, and the transitions over the potential
barriers are triggered by  thermal activation. The limiting
force thresholds in this case read: $F_1 \simeq -4.86$ and $F_2
\simeq 5.5$.

At zero bias ($F=0$), the quantum and classical motors now proceed
in the {\it opposite directions}  within a large range of
temperatures, see in the central panel in Fig. \ref{fig2}. By
applying a large enough constant load either into  positive or 
negative direction, this feature is seemingly destroyed. The motors
are forced to transport accordingly to the applied bias. For  very
small values of the force, however, this very intriguing behavior
induced by quantum fluctuations is still preserved at low
temperatures.

\begin{figure}[htbp]
  \begin{center}
    \includegraphics[angle=0,width=0.9\linewidth,clip=]{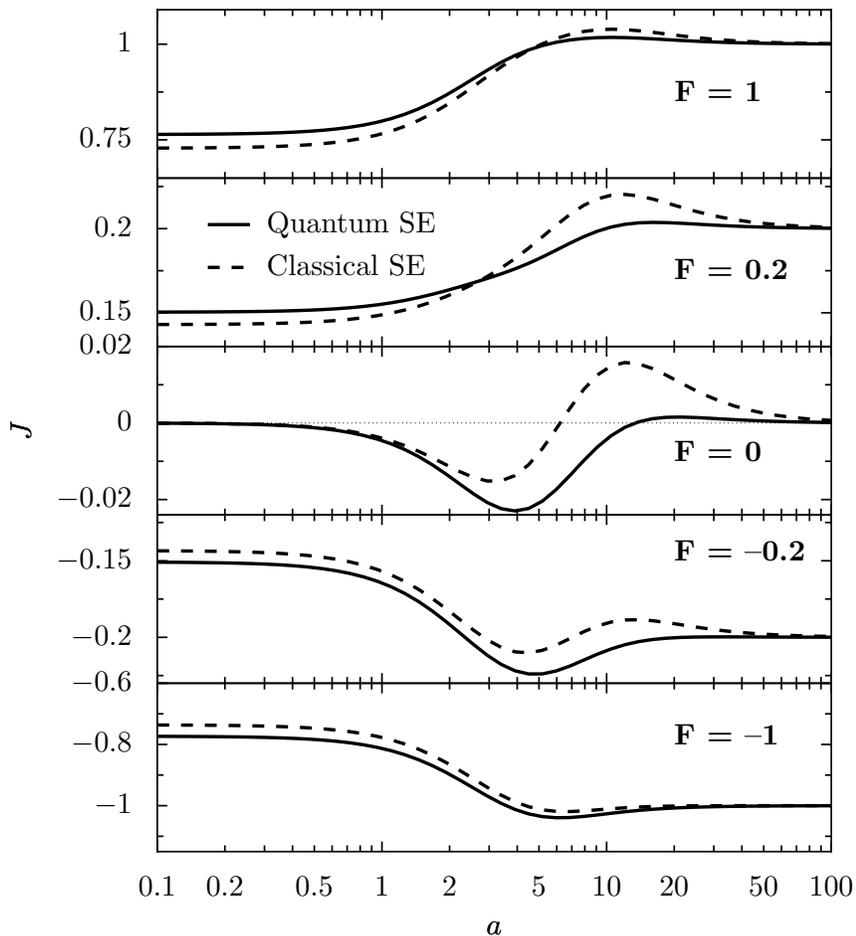}
  \end{center}
  \caption{The average current $J$ of the quantum
  (solid line) and
  classical (dashed line) Brownian motor versus the two-state
  noise amplitude $a$ is shown  in order to elucidate
  the influence of quantum fluctuations on the directed transport.
  We present the current for several values of constant load force
  $F=-1,-0.2,0,0.2,1$, from bottom toward top.
  The dimensionless inverse temperature reads $\beta_0 = 2$ and the
  potential
  shape parameter
  $B$ is set to $0.3$, cf.  panel (a)
  in Fig \ref{fig0}.
  }
  \label{fig3}
\end{figure}

In the Fig. \ref{fig3} we plot the current as a function of the
dichotomic noise amplitude  $a$. The ratchet potential  is the same
as the one used in  Fig.  \ref{fig1}. The rescaled inverse
temperature is set rather large at $\beta_0=2$. We compare the
resulting  classical and quantum currents for five various  values
of the constant force $F=-1, -0.2, 0, 0.2, 1$.

The middle panel of Fig. \ref{fig3} depicts the Brownian motor
currents for the unbiased case  $F=0$. We recover a distinct feature
of the ratchet dynamics, namely, the occurrence of current reversals
\cite{KosLuc2001}, both in the classical and in the quantum case,
located however at different $a$-values. In presence of a
sufficiently large bias force these current reversals disappear
and  the classical and quantum currents approach each other.

For small values of the dichotomic noise strength $a$ and
in the absence of the external load, classical and quantum currents
assume almost 
identical values, see the central panel in Fig. 4. 
In this regime of  small dichotomic noise strengths
and non-zero bias, the absolute value of the 
quantum current is always larger than its classical counterpart.

\section{Summary}

Quantum noise induced, directed transport features of an overdamped
Brownian motor moving in a spatially periodic ratchet potential,
that is exposed to the constant load in the presence of
nonequilibrium, adiabatically varying dichotomic fluctuations, are
investigated with this work. The classical and the quantum regimes,
being determined by the ratio of two energy scales (the energy of
thermal equilibrium fluctuations and the activation energy over
barriers), are analyzed in detail. There is no a general rule on 
the influence of quantum corrections on current: as a function of the system
parameters, the quantum effects may either increase or as well
decrease the average current of the so forced Brownian motors.
The impact of quantum corrections is also  clearly encoded
in a shift of position of the respective values for the stall force.

\section*{Acknowledgments}
This  work has been supported  by the ESF programme: Stochastic
Dynamics, the Polish Ministry of Science and Higher Education via
the grant No PBZ-MIN-008/P03/2003.


\begin{thebibliography}{99}

\bibitem{chaos}
P. H\"anggi and F. Marchesoni, Chaos {\bf 15}, 026101 (2005).

\bibitem{clas1}
P. Reimann and P.~H\"anggi, Appl.~Phys. A {\bf 75}, 169 (2002); \\
P. Reimann, Phys. Rep. {\bf 361}, 57 (2002).

\bibitem{clas2}
R. D. Astumian and P. H\"anggi, Physics Today {\bf 55} (No. 11), 33
(2002).

\bibitem{clas3}
P. H\"anggi, F. Marchesoni, and F. Nori, Ann. Phys. (Berlin) {\bf
14}, 51 (2005).
%
\bibitem{quant1}
P. Reimann, M. Grifoni, and P. H\"anggi , Phys. Rev. Lett. {\bf 79},
10 (1997);\\ P. Reimann and P. H\"anggi, Chaos {\bf 8}, 629 (1998).
%
\bibitem{quant2}
I. Goychuk, M. Grifoni,  and P. H\"anggi , Phys. Rev. Lett. {\bf
81}, 649 (1998);\\ I. Goychuk, M. Grifoni,  and P. H\"anggi, Phys.
Rev. Lett. {\bf 81}, 2837 (1998) (erratum);\\ I. Goychuk and P.
H\"anggi, Europhys. Lett. {\bf 43}, 503 (1998);\\ M. Grifoni,
M.S.~Ferreira,
J. Peguiron, J.B.~Majer, Phys. Rev. Lett. {\bf 89}, 146801 (2002);\\
S. Scheidl, and V.M. Vinokur, Phys. Rev. B {\bf 65}, 195305 (2002).

\bibitem{MacKos2004}
  L. Machura, M. Kostur, P. H\"anggi, P. Talkner, and J. \L uczka, Phys. Rev. E {\bf 70}, 031107 (2004).

\bibitem{MacKos2006}
  L. Machura, M. Kostur, P. Talkner, J. \L uczka, and P. H\"anggi, Phys. Rev. E {\bf 73}, 031105 (2006).

\bibitem{ingoldchaos}
P.~H\"anggi and G.~L.~Ingold, Chaos {\bf 15}, 026105 (2005).

\bibitem{anker}
  J. Ankerhold, P. Pechukas, and H. Grabert, Phys. Rev. Lett. {\bf 87}, 086802
  (2001);\\
  J. Ankerhold, Phys. Rev. E {\bf 64}, 060102 (2001).


\bibitem{ingold85}
P. H\"anggi, H. Grabert, G. L. Ingold and U. Weiss, Phys. Rev. Lett.
{\bf 55}, 761 (1985).

\bibitem{LucRud2005} J. \L uczka, R. Rudnicki, P. H\"anggi, Physica A {\bf 351}, 60 (2005).

%
\bibitem{HT} P. H\"anggi and H. Thomas, Phys. Rep. {\bf 88}, 207 (1982).
%
%
\bibitem{kul} J. Kula, T. Czernik, J. {\L}uczka, Phys. Rev. Lett. {\bf 80}, 1377
(1998).

\bibitem{KosLuc2001}
  M. Kostur, and J. \L uczka, Phys. Rev. E {\bf 63}, 021101 (2001).

\end{thebibliography}
\end{document}